\documentclass[twocolumn,superscriptaddress,showpacs,aps,prx]{revtex4-1}
\usepackage{amsfonts}
\usepackage{amsmath}
\usepackage{amssymb}
\usepackage{bm}
\usepackage{graphicx}
\usepackage{txfonts}
\usepackage{tabularx}
\usepackage{multirow}
\usepackage{subfigure}
\usepackage[colorlinks,linkcolor=blue,anchorcolor=blue,citecolor=blue,urlcolor=blue,dvipdfm]{hyperref}%
\makeatletter

\newcommand{\Rmnum}[1]{\expandafter\@slowromancap\romannumeral #1@}
\makeatother
\begin{document}
\title{Few-Shot Machine Learning in 3D Ising Model}
\author{Rui Zhang}
\thanks{These two authors contributed equally to this work.}
\affiliation{SKLSM, Institute of Semiconductors, Chinese Academy of Sciences, 100083 Beijing, China}
\affiliation{Center for Excellence in Topological Quantum Computation, University of Chinese Academy of Sciences, Beijing 100190, China}
\affiliation{COT Design Department, HiSilicon Technologies Co., LTD, 518129 Shenzhen, China}
\author{Bin Wei}
\thanks{These two authors contributed equally to this work.}
\affiliation{SKLSM, Institute of Semiconductors, Chinese Academy of Sciences, 100083 Beijing, China}
\affiliation{Center for Excellence in Topological Quantum Computation, University of Chinese Academy of Sciences, Beijing 100190, China}
\author{Dong Zhang}
\affiliation{SKLSM, Institute of Semiconductors, Chinese Academy of Sciences, 100083 Beijing, China}
\affiliation{Center for Excellence in Topological Quantum Computation, University of Chinese Academy of Sciences, Beijing 100190, China}
\author{Jia-Ji Zhu}
\email[Corresponding author: ]{zhujj@cqupt.edu.cn}
\affiliation{Institute for Quantum Information and Spintronics, School of Science, Chongqing University of Posts and Telecommunications, 400061 Chongqing, China}
\author{Kai Chang}
\email[Corresponding author: ]{kchang@semi.ac.cn}
\affiliation{SKLSM, Institute of Semiconductors, Chinese Academy of Sciences, 100083 Beijing, China}
\affiliation{Center for Excellence in Topological Quantum Computation, University of Chinese Academy of Sciences, Beijing 100190, China}

\begin{abstract}
We investigate theoretically the phase transition in three dimensional cubic Ising model utilizing state-of-the-art machine learning algorithms. Supervised machine learning models show high accuracies (~99\%) in phase classification and very small relative errors ($< 10^{-4}$) of the energies in different spin configurations. Unsupervised machine learning models are introduced to study the spin configuration reconstructions and reductions, and the phases of reconstructed spin configurations can be accurately classified by a linear logistic algorithm. Based on the comparison between various machine learning models, we develop a few-shot strategy to predict phase transitions in larger lattices from trained sample in smaller lattices. The few-shot machine learning strategy for three dimensional(3D) Ising model enable us to study 3D ising model efficiently and provides a new integrated and highly accurate approach to other spin models.

\end{abstract}
\maketitle

\section{Introduction}
Nowadays, the machine learning techniques have achieved remarkable progresses in many different fields, \textit{e.g.}, image recognition, fraud detection, natural language processing, auto driving, \textit{etc.}, due to their powerful magic to extract features from huge data sets without explicit guidance from a human programmer. In physics, machine learning algorithms have been employed to study many-body physics\cite{carleo,arsenault,caizi}, strong correlated systems\cite{tomoki,kelvin}, electronic or transport properties\cite{faber,rupp,lopez}, structure predictions\cite{jacobsen,schoenholz,montavon,isayev}, phase matter or phase transitions\cite{torlai,Carra,zhangyi,broecker,torlai1,ponte}, and different spin models\cite{Carra,liujunwei,wanglei,huwenjian,wangce,wetzel}.

Among the various studied spin models, the Ising model attracts the most interests, because it is not only rich in physics and mathematics, such as phase transitions\cite{lee} and Kolmogorov's zero-one laws\cite{kolmogorov}, but also inspirations in many other fields, such as social sciences\cite{stauffer2008social} and neuroscience\cite{schneidman2006weak}. The possible applications of the Ising model are extremely large. For instance, it can be used for magnetic insulators\cite{cowley1972properties}, binary alloys\cite{binder1974kinetic,richards1971pairwise,clapp1968correlation}, lattice gas model for fluids\cite{binder1982monte}, ferroelectrics\cite{blinc1987proton}, biological systems\cite{hopfield1982neural} or general demonstration of statistical mechanics\cite{baxter2016exactly}.

Of particular interest is the three-dimensional (3D) Ising model, which has been introduced to exactly describe the second generation D-Wave quantum computer of 512 qubits. Very recently, the quantum supremacy makes the 3D Ising model again a hot topic\cite{boixo2018characterizing}. Although the Ising model looks very simple, the complexity of the solutions to the model increases along with the dimensions. Till now, the Ising model has only been solved mathematically rigorously in one \cite{ising1925beitrag} and two dimensional lattices\cite{onsager1944crystal}. In three dimension or higher dimensional case, the analytic solutions remain unavailable. Instead, various approximation methods have been developed to study ising model as a last resort, including the Conformal Bootstrap approach\cite{poland2016,el2014,el2012}, numerical approaches, such as Monte Carlo simulation\cite{blote} and mean field theories\cite{jensen1983mean}.

It is notable that there exists a reciprocal relationship between the Ising model and neural networks or machine learning algorithms. On the one hand, the Ising model has played a critical role in the development of machine learning techniques. For example, the famous Hopfield network\cite{hopfield1982neural} is a typical recurrent artificial network based on dynamical Ising model. On the other hand, the machine learning techniques provide a new effective approach to solve the Ising model: J. Carrasquilla \textit{et al.} have investigated the phase transition of 2D Ising model recently, utilizing supervised learning models such as neural networks (NNs) and 2D convolutional neural networks (CNNs)\cite{Carra}, and demonstrated the machine learning techniques can study the phase transition of 2D Ising model with very high accuracy; Besides the supervised learning models, unsupervised learning models such as principal component analysis (PCA) and auto encoder (AE) have also been applied to study 2D XY model\cite{huwenjian,wangce} and 3D XY model\cite{wetzel}. However, the 3D Ising model still remains unexplored using the machine learning techniques.

In this work, we have studied the phase transition of ferromagnetic 3D Ising model by Monte Carlo-sampled machine learning techniques. Instead of standard Monte Carlo simulations\cite{OLFF1989379,PREIS20094468}, the needed parameters are predicted by the state-of-the-art machine learning techniques. In order to compare accuracy and efficiency of current machine learning algorithms, we perform both supervised and unsupervised machine learning algorithms in the Ising models. Since the computation consumptions increase sharply with increasing the lattice sizes, we develop a few-shot machine learning strategy by combining specific supervised and unsupervised machine learning algorithms, to predict phase transitions in larger lattices from trained data in smaller lattices. The paper is organized as follows: In Sec. \Rmnum{2}, we present the 3D Ising model and methods used during the calculations. In Sec. \Rmnum{3} and Sec. \Rmnum{4}, we demonstrate the performances of machine learning algorithms and the few-shot machine learning strategy.
In Sec. \Rmnum{5}, we report the conclusions of our work.

\section{Models and Methods}

The Hamiltonian of the ferromagnetic 3D Ising model in a cubic lattice is given by:
\begin{equation}
H=-J\sum_{\langle i,j\rangle }\sigma _{i}\sigma _{j},  \label{eq:ising}
\end{equation}%
with uniform interaction strength $J=1$ and a binary spin configurations $\sigma _{i}\in \{+1=\uparrow ,-1=\downarrow \}$ on each site $i$ where $i$ runs over all the spin sites ($N$). The sum goes over all nearest-neighbor pairs and the periodic boundary condition is taken in our calculation. The magnetization of a spin sample is defined as $M=\frac{1}{N}\sum_{i}\sigma_{i}$. In the 3D Ising model under studied, there are $2^{N}$ different spin configurations for cubic lattices, and two key physical parameters, the transition temperature ($T_{c}$) and the critical exponent ($v$).

\begin{figure}[h]
\centering
\includegraphics[width=0.99\columnwidth]{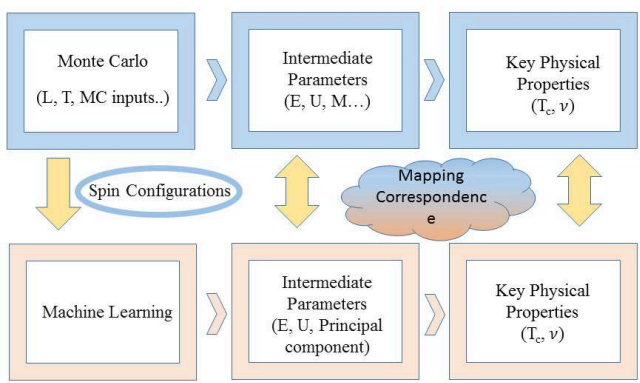}
\caption{The upper panel: the standard Monte Carlo procedure to capture the key physical parameters $T_{c}$ and $v$; The lower panel: Monte Carlo based machine learning procedure. The relationship between the MC and the ML procedures are indicated by the arrows in the middle. }
\label{fig:procedure}
\end{figure}

To investigate the two key physical parameters, we consider cubic lattices with different sizes ($L$) and adopt a Monte Carlo-sampled machine learning procedure as illustrated in Fig.\ref{fig:procedure}. In Fig.\ref{fig:procedure}, the upper panel indicates a standard Monte Carlo procedure, in which all the intermediate parameters, such as energy (E), binder cumulant (U), magnetization (M), etc., for fitting the key physical quantities $T_{c}$ and $v$, are calculated by standard Monte Carlo algorithm itself. While in the Monte Carlo-sampled machine learning procedure as illustrated in the lower panel, the Monte Carlo method is applied to generate raw data sets to feed into machine learning models. At the first stage, the input data sets will be minimized to spin configurations only. Afterwards, we prove the exact correspondence between the intermediate parameters given by standard Monte Carlo simulation and machine learning models. Finally, one can compare the predicted $T_{c}$ and $v$ from both procedures.

The Monte Carlo simulation, implemented by the Swendsen-Wang cluster algorithms (SWCA)\cite{Swendsen}, is performed to calculate the transition temperatures ($T_{c}$) by creating 2000 independent spin configurations at each temperature step, and to provide initial raw data sets to train the machine learning models. The SWCA algorithm updates a cluster of spins for each step, which reduces the correlation of data remarkably. We take 50, 000 update steps between every two output of spin configuration in order to reduce the data correlation more efficiently.

Machine learning methods are applied to get above parameters and are categorized as supervised and unsupervised learning models as follows. Supervised learning is the machine learning task of inferring a function from labeled training data\cite{Mehryar}, in which each sample is a pair of an input object and its desired label. A supervised learning algorithm learns from the training sets and then produces an inferred function which can be applied to the test sets that have never been trained before. The accuracy of the supervised learning model is determined by comparing the output of the trained model and the original labels on the test sets, which is also an important benchmark to evaluate the performance of the machine learning model. In this case, the input data for the supervised learning models are the spin configurations of the cubic Ising lattice. The corresponding labels are determined by the corresponding $T$ of the input configurations: the spin configurations are labeled by (1, 0) when $T<T_{c}$, and by (0, 1) when $T>T_{c}$. Three supervised learning models are employed in this work: support vector machine (SVM)\cite{svm,svmkernel}, neural network (NN) and 3D convolution neural network (CNN)\cite{ann,hinton}.

Contrary to supervised learning method, unsupervised learning models are expected to infer a function to describe hidden structure from unlabeled data which can be a classification or categorization not included in the given physical observability. To solve complicated physical problem, the labels of target system are usually difficult to get from calculations or experiments, which makes the unsupervised machine learning necessary and useful. Since the input data sets to the unsupervised learning are merely spin configurations, the spin configurations with different labels will fall into different clusters as the unsupervised learning models are well trained. In this work we take two commonly-used unsupervised machine learning model, the principal component analysis (PCA)\cite{pca,pca1} and the restricted Boltzmann machine (RBM) . All the supervised and unsupervised machine learning models are trained by the spin configurations generated with $T\in \lbrack 3,6]$ in units of $J/k$ in total $n=100,000$ configurations for every size.

\section{Conventional machine learning in 3D Ising model}

First, we perform supervised classification for spin configurations in 3D Ising model utilizing SVM, NN and 3D CNN algorithms. The input data for supervised machine learning models are obtained from the Monte Carlo sampling at different temperatures $T$ in a finite cubic lattice $L=16$. The spin configurations are binary-labeled by their corresponding phases, $i.e.$, high temperature paramagnetic phase when $T>T_{c}$ and the low temperature ferromagnetic phase when $T<T_{c}$ \cite{Carra}. The data sets are randomly divided into the train and test sets, all the supervised machine learning models are trained in the the train sets and predicted in the test sets. In this work, we adopt SVM model which is implemented in Scikit-Learn library\cite{sklearn,sklearn_api} with optimized hyper parameters, and build the NN and 3D CNN models in the framework of Keras library\cite{keras}.
\begin{figure}[h]
\centering
\includegraphics[width=0.99\columnwidth]{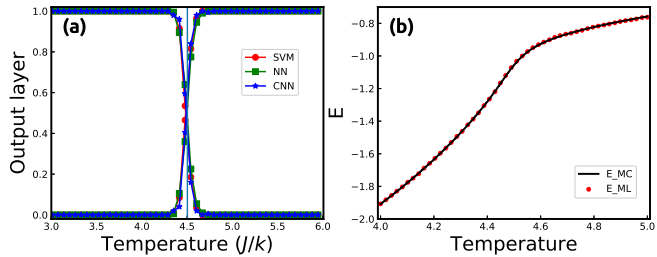}
\caption{(a) The average outputs of different ML models SVM (red dot), NN (green rectangle), and 3D CNN (blue star) on the test set with $L=16$. (b) The average energies predicted by 3D CNN model, the black solid line and the red dots are obtained from the Monte Carlo simulations and 3D CNN model, respectively. }
\label{fig:supervised}
\end{figure}

The numerical results obtained from the SVM, NN and 3D CNN models are shown in Fig. \ref{fig:supervised}(a). All the supervised machine learning models can classify the high and low-temperature phases with accuracies about 99\%, and the average outputs of the supervised machine learning models cross at the predicted $T_{c}$ which is very close to the Monte Carlo simulations. The result shows that all the three supervised machine learning models used in our calculations are effective to distinguish the high and low-temperature phases, although they are quite different.

Machine learning models can not only classify phases of 3D Ising model, but also predict physical quantities, such as the transition temperature $T_{c}$, the critical exponents $\nu$ and the averaged energies of various spin configurations. We train a 3D CNN model to predict energies of different spin configurations (see Fig. 2(b)). The predicted energies agree very well with the Monte Carlo simulations, the relative error of the predicted energy is smaller than $1.0\times 10^{-4}$.

Till now, all the input data are directly used to train the supervised machine learning models without any pre-processing. Next, we do some pre-processing to the 3D spin configurations. Two unsupervised machine learning models, PCA and RBM, are considered to reduce dimensions and reconstruct 3D spin configurations, respectively. We take the PCA and Bernoulli RBM implemented in Scikit-Learn library\cite{sklearn,sklearn_api}.
\begin{figure}[h]
\centering
\includegraphics[width=0.99\columnwidth]{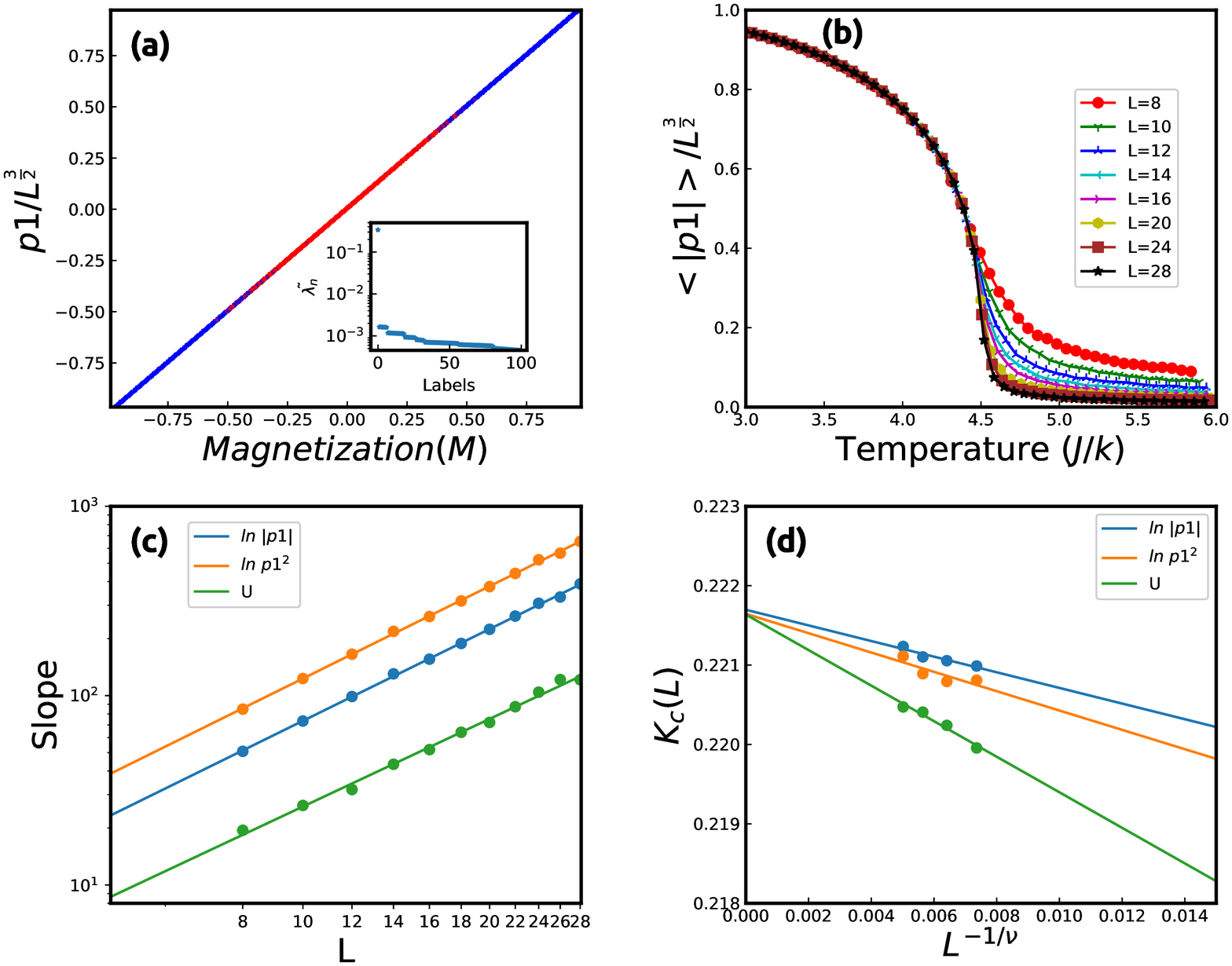}
\caption{(a) The output of the first principal component as a function of the magnetization ($M$) in a cubic lattice with $L=16$. The explained variance, $i.e.$, the ratio between the variance of the principal components and the total variance, are shown in the inset of (a). (b) Absolute value of the first PCA output as a function of temperature. (c) Finite size scaling analysis to determine critical exponent $v$ based on the first principal component and energy, the critical exponent is fitted to be 0.629. (d) Finite size scaling analysis to determine transition $K_{c}$. }
\label{fig:unsupervised}
\end{figure}

We consider one principal component when we apply PCA to 3D spin configurations, since we find that the first principal component possesses the greatest explained variance (see the inset of Fig. \ref{fig:unsupervised}(a)). In Fig. \ref{fig:unsupervised}(a), one can find that the outputs $p1$ are linear with the magnetization $M$ , which means the correlation between $p1$ and $M$ is almost 1.0 and the physical observable, magnetization, can be learned from the first PCA of the input data. We also plot the absolute value of $p1$ as a function of temperature in Fig. \ref{fig:unsupervised}(b), and find that the lines with different size parameter $L$ tend to approach together when the lattice size $L$ increases. Therefore we take finite size scaling analysis to determine the critical exponent $v$ and predict the transition temperature of 3D Ising model with infinite $L$ based on the first principal component and the predicted energy\cite{Ferrenberg081,Andrea549}.

The results of finite size analysis are shown in Fig. \ref{fig:unsupervised}(c) and Fig. \ref{fig:unsupervised}(d). The critical exponent of 3D Ising model is a very important physical quantity to describe the critical behavior of the phase transition, and still has not been studied utilizing the machine learning models\cite{huwenjian,wangce,wetzel,PhysRevE.97.032119,Wang2018}. Here it can be determined based on cubic spin configurations with $L\in \lbrack 8,28]$. $K_{c}=\frac{1}{T_{c}}$ is obtained by fitting the data with $L\geq 22$. The critical exponent $v$ and the critical temperature $T_{c}$ are measured to be 0.629 and 4.511417 ($K_{c}=0.2216598$), respectively, which agree very well with the results obtained from the renormalization group theory\cite{Ferrenberg081,Andrea549}.

\begin{figure}[h]
\centering
\subfigure{\includegraphics[width=0.49\columnwidth]{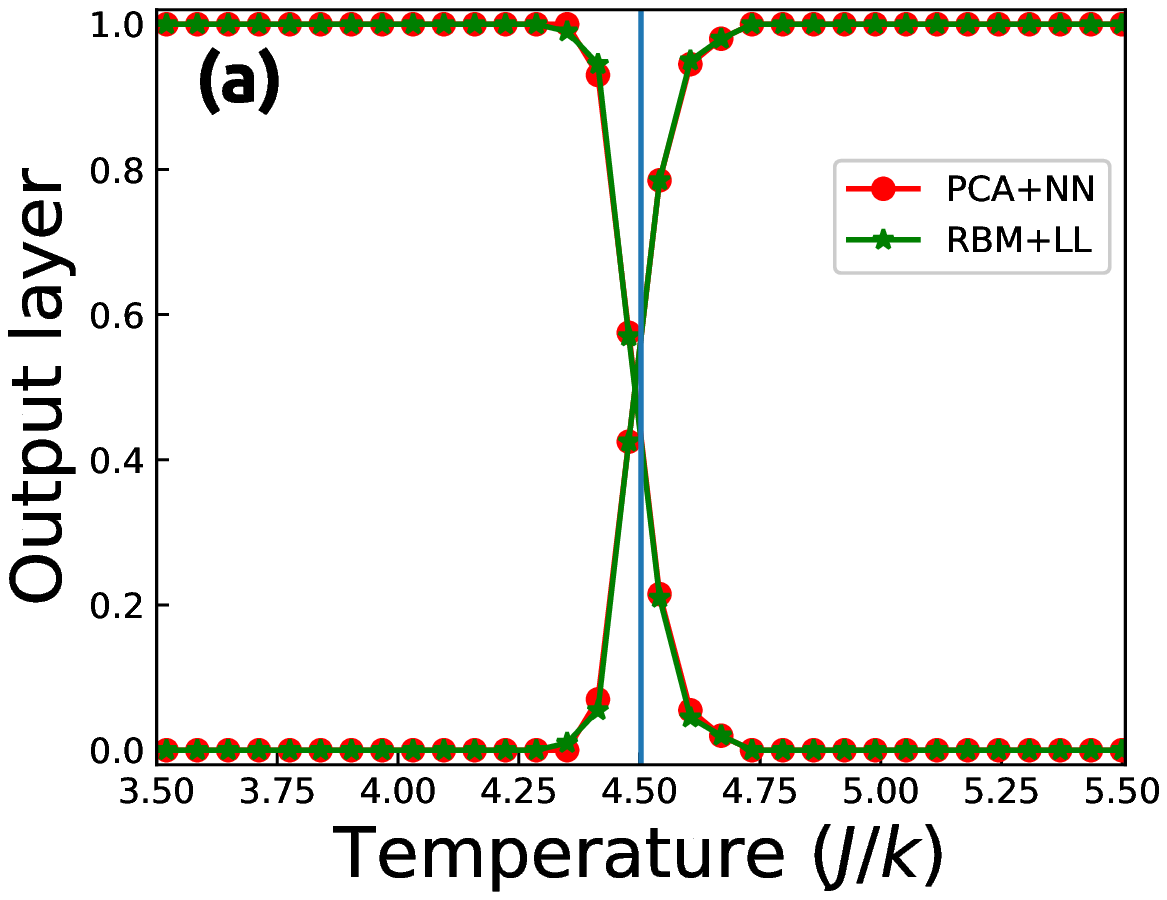}%
\label{fig:rbm_pca}} \subfigure{\includegraphics[width=0.49%
\columnwidth]{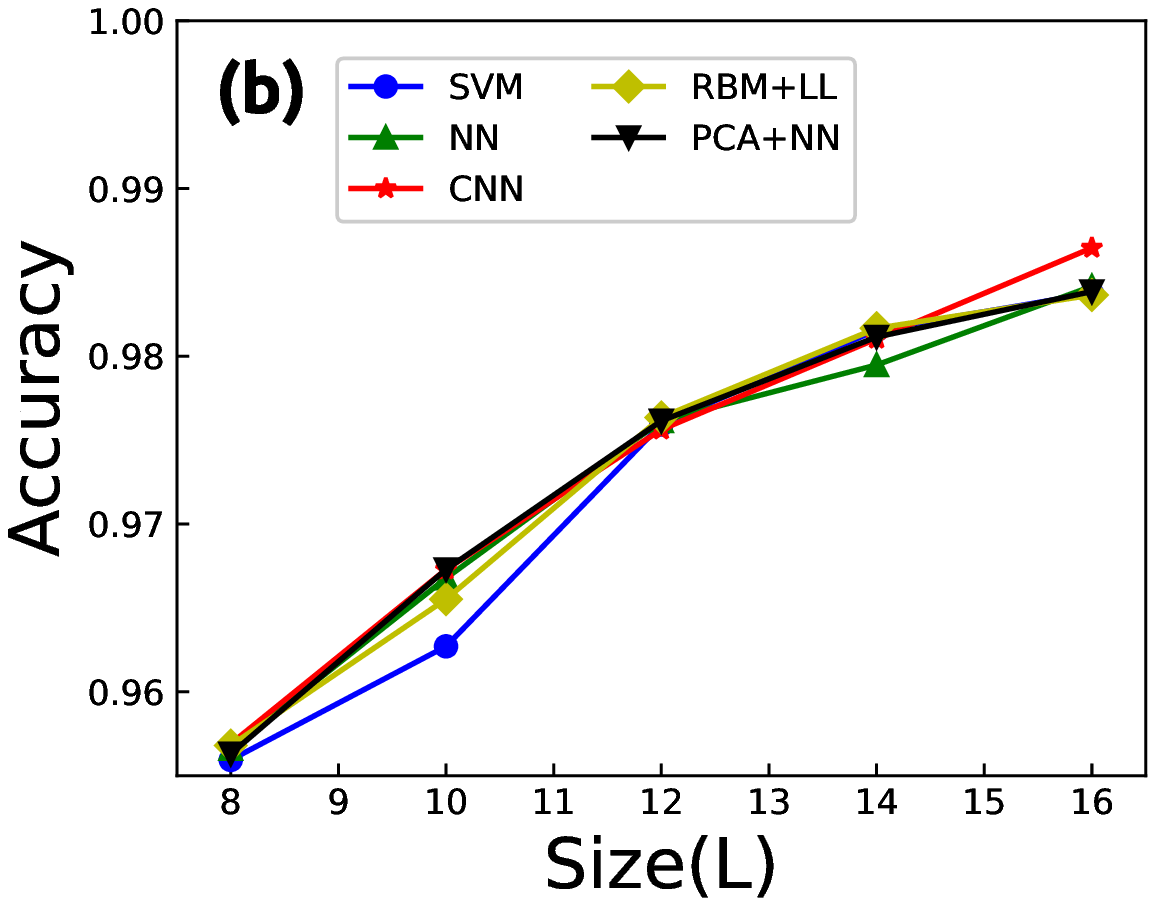}\label{fig:acc_dim}}\newline
\caption{(a) The average outputs of different RBM combined with a linear logistic classifier, and PCA combined with a NN classifier, both with $L=16$. (b) Accuracies as a function of $L$ for different supervised and combined supervised models. }
\label{fig:combination}
\end{figure}

Since $p1$ is also closely related to the high- and low-temperature phases (see the red and blue scatters shown in Fig. \ref{fig:unsupervised}(a)), this feature makes it possible to classify different phases based on the PCA outputs. In addition, the RBM is applied to reconstruct 3D spin configurations. The results obtained from supervised learning models based on the reduced and reconstructed spin configurations are shown in Fig. \ref{fig:combination}(a). From this figure, one can see clearly that, the combined model can be used to distinguish different phases with enough high accuracy. Notice that the RBM reconstructed 3D Ising model can be simply classified by a logistic linear classifier. The reconstrution process differentiates various spin configurations, and the PCA reduction process makes the classification procedure much more efficient than previous direct classification machine learning models.

A full comparison between pre-processed machine learning models and direct supervised machine learning models is shown in Fig. \ref{fig:combination}(b). One can find that the overall accuracies increase with the size parameter $L$ and the performance of all these machine learning models approach each other. The different behaviors of these models arise from the complexity of these machine learning models, and the pre-processed model could require less computation resources but without loss of accuracies.

\section{Few-shot machine Learning in 3D Ising Model}

Supervised machine learning models trained from the data with the lattice size $L$ cannot be applied to the lattices with $L+\Delta L$ directly. In order to make it possible to predict the phase transition in a larger size $L+\Delta L$, we combine the PCA and the NN together. We first perform PCA to the spin lattice with different sizes (e.g. L=10,12,14,16,20,24), only the first leading two principal components are remained as the input for next step.

\begin{figure}[h]
\centering
\includegraphics[width=0.99\columnwidth]{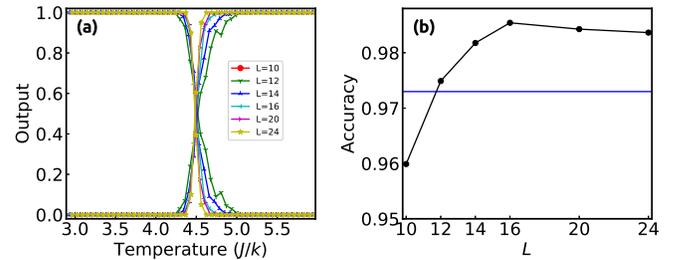}
\caption{The performance of the model trained by mixing data with different $L$. (a) The average output of the mixed trained model as a function of temperature. (b) The overall accuracies of the mixed trained model tested on different $L$.}
\label{fig:mix}
\end{figure}

Such a bunch of pre-processed data makes next-step train sets. Then the train sets are fed to a NN model which consists of an input layer with two units, one hidden layer with 30 units, and at the end the output layer with 2 units. The performances of the mixed trained model on test sets are shown in Fig. \ref{fig:mix}(a), and one can find that the mixed trained model performs very good with different size parameters. The overall accuracies of this mixed trained model obtained from the test sets with different $L$ are shown in Fig. \ref{fig:mix}(b), and the performance of the mixed trained model is comparable to the models based on particular $L$ shown in Fig. \ref{fig:acc_dim}.
\par
\begin{figure}[h]
\centering
\includegraphics[width=0.99\columnwidth]{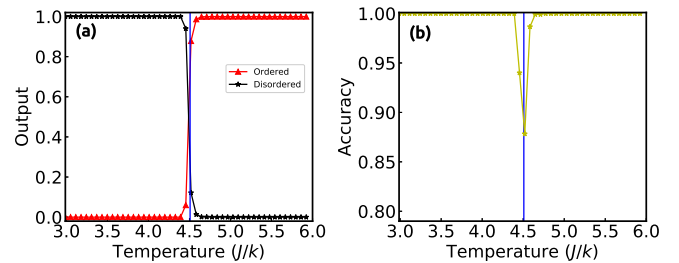}
\caption{The performance of the mixed trained model on the unexplored size $L=28$. (a) The average output of the mixed trained model as a function of temperature. (b) The accuracies of the model at different temperatures $T$. }
\label{fig:extra}
\end{figure}

After the confirmation of the validity of the mixed trained model for 3D Ising models in known size lattices, we extend it to unexplored size lattice. By applying the mixed trained model to the spin configurations with $L=28$, we find out that the model can predict the $L=28$ lattice with an amazing accuracy $0.996$, therefore verify the capability of the model to predict critical temperatures in larger lattices which have never been trained. The average outputs at different temperatures are shown in Fig. \ref{fig:extra}(a), and the predicted phase transition temperature is very close to the standard Monte Carlo simulation (the blue vertical line in Fig. \ref{fig:extra}(a)). Fig. \ref{fig:extra}(b) shows that the mixed trained model possesses very high accuracies at different temperatures $T$, even when $T$ is very close to the critical temperature $T_{c}$. Since the prediction of larger lattice can be obtained merely from training smaller samples, the combination of PCA and NN enables us to predict the phase transition in a quite large lattice which is unattainable by standard Monte Carlo procedures. This mixed trained model demonstrates a convincible few-shot machine learning strategy for spin models.

\section{Conclusions}
In conclusion, we perform several machine learning models for 3D Ising models. Based on the state-of-the-art Monte Carlo simulations, we find both supervised and unsupervised machine learning models can extract physical information from raw data. A critical exponent and the transition temperature are accurately fitted by PCA. The combination of RBM and a simple linear logistic classifier is proven to be as powerful as supervised machine learning models for 3D Ising models. We show a new type of few-shot machine learning strategy based on PCA and a feed forward NN classifier. The few-shot machine learning strategy for 3D Ising model makes it possible to predict the phase transition in a larger lattice based on data from some smaller samples, which also provides an approach to other spin models. Our work presents an integrated and highly accurate machine-learning toolbox and demonstrates a new few-shot learning strategy for studying more spin models.
\section{ACKNOWLEDGMENTS}
This work was supported by NBRPC Grant No. 2016YFE0110000, 2015CB92150 and NSFC Grants No. 11504336 and 11404043.

\section*{APPENDIX}
\section*{A: ARCHITECTURES OF NEURAL NETWORKS}
In this work we take different deep learning (DL) models, here we give the architectures of full connected neural networks (NN) and 3D convolutional neural networks(3D CNN). The full connected NN model consists of four kinds of layers: input layer, hidden layer, dropout layer, and output layer. The input layer contains $L^3$ neurons, the hidden layer contains 100 neurons which are activated by sigmoid function, the output layer contains two neurons which are activated by softmax functions. We take a dropout rate 0.1 to prevent overfitting of this model.

The 3D CNN model is composed of a 3D input layer, 3D convolutional layers, a dropout layer, a full connected hidden layer, and the output layer.  The 3D convolutional layers apply 64 2 x 2 filters to the spin configurations, the full connected hidden layer contains 64 neurons. All the layers are activated by sigmoid functions except for the output layer which is activated by softmax function. The dropout layer is added between the 3D convolutional layers and the output layer with dropout rate 0.25.
\section*{B: ALGORITHM COMPLEXITY}
The mixed-trained model consists of two procedures: (1) applying PCA to all the spin configurations of train set and (2) applying NN to the reduced spin configurations. The determine complexity of this model comes from the PCA step. In this paper we implement PCA utilizing Scikit-Learn with randomized truncated singular value decomposition (SVD)\citep{halko,keras},
the complexity is $O(n_{max}^{2}\cdot n_{components})$. Here $n_{max}$ is the max value between the number of spin configurations ($n_{samples}$) and dimensions of the spin configurations($L^{3}$), $n_{components}$ is the number of principal components\citep{halko,keras}. When the size of 3D Ising model is big enougth ($L^{3}>n_{samples}$), the complexity is propotional to $L^{6}$ which grows rapidly when L increase.

\bibliographystyle{apsrev4-1}
\bibliography{ising}
\end{document}